\documentclass[12pt,preprint]{aastex}
\usepackage{psfig}
 
\usepackage{graphicx}
\usepackage{epstopdf}
\usepackage{psfig}

\def\deg      {{\ifmmode^\circ\else$^\circ$\fi}} 

\slugcomment{To appear in the ApJ letters}
 
\shorttitle{Ionization near-zones associated with quasars at 
$z\sim 6$}
\shortauthors{Carilli et al.}
 
\begin{document}
\title{Ionization near-zones associated with quasars at  $z\sim 6$}

\author{C.L. Carilli\altaffilmark{1}, Ran Wang\altaffilmark{1,2}, 
   X. Fan\altaffilmark{3},  
  F. Walter\altaffilmark{4}, J. Kurk\altaffilmark{4}, 
  D. Riechers\altaffilmark{5},
  J. Wagg\altaffilmark{6},
  J. Hennawi\altaffilmark{4},  L. Jiang\altaffilmark{3},
  K.M.Menten\altaffilmark{7}, F. Bertoldi\altaffilmark{8}, Michael A. 
  Strauss\altaffilmark{9}, P. Cox\altaffilmark{10} }

\email{ccarilli@aoc.nrao.edu}

\altaffiltext{$\star$}{The Very Large Array of the National Radio Astronomy
Observatory, is a facility of the National Science Foundation
operated under cooperative agreement by Associated Universities, Inc}

\altaffiltext{1}{National Radio Astronomy Observatory, P. O. Box 0,
Socorro, NM 87801}

\altaffiltext{2}{Purple Mountain Observatory, CAS 2, West Beijing Road,
Nanjing, China, 210008}

\altaffiltext{3}{Steward Observatory, University of Arizona,
Tucson, AZ, USA, 85721}

\altaffiltext{4}{Max-Planck Institute for Astronomy, K\"onigstuhl 17, 69117,
Heidelberg, Germany}
                                                                               
\altaffiltext{5}{Department of Astronomy, California Institute of Technology,
MC 249-17, 1200 East California Boulevard, Pasadena, CA 91125, USA; Hubble
Fellow}

\altaffiltext{6}{ESO-ALMA, Alonso de Cordova 3107 Vitacura Casilla
19001 Santiago 19, Chile}

\altaffiltext{7}{Max-Planck Institute for Radio Astronomy, Auf dem H\"ugel 49,
Bonn, Germany}
                                                                             
\altaffiltext{8}{Argelander Institute for Astronomy, University of
Bonn, Auf dem H\"ugel 71, 53121 Bonn, Germany}
                         
\altaffiltext{9}{Princeton University Observatory, Peyton Hall,
Princeton, NJ 08544, USA}
                                                    
\altaffiltext{10}{Institut de Radioastronomie Millimetrique, 
300 rue de la Piscine, 38406 Saint Martin d'Heres, France}

\begin{abstract}
\noindent 

We analyze the size evolution of HII regions around 27 quasars between
$z=5.7$ to 6.4 ('quasar near-zones' or NZ). We include more sources
than previous studies, and we use more accurate redshifts for the host
galaxies, with 8 CO molecular line redshifts and 9 MgII redshifts.  We
confirm the trend for an increase in NZ size with decreasing redshift,
with the luminosity normalized proper size evolving as: $\rm R_{NZ,corrected}
= (7.4 \pm 0.3) - (8.0 \pm 1.1) \times (z-6)$ Mpc.  While
derivation of the absolute neutral fraction remains difficult with
this technique, the evolution of the NZ sizes suggests a decrease in
the neutral fraction of intergalactic hydrogen by a factor $\sim 9.4$
from $z=6.4$ to 5.7, in its simplest interpretation.  Alternatively,
recent numerical simulations suggest that this rapid increase in
near-zone size from $z=6.4$ to 5.7 is due to the rapid increase in the
background photo-ionization rate at the end of the percolation or
overlap phase, when the average mean free path of ionizing photons
increases dramatically.  In either case, the results are consistent
with the idea that $z \sim 6$ to 7 corresponds to the tail end of
cosmic reionization. The scatter in the normalized NZ sizes is larger
than expected simply from measurement errors, and likely reflects
intrinsic differences in the quasars or their environments. We find
that the near-zone sizes increase with quasar UV luminosity, as
expected for photo-ionization dominated by quasar radiation.

\end{abstract}


\section{Introduction}

Observations have set the first constraints on the epoch of
reionization (EoR), corresponding to the formation epoch of the first
luminous objects (see the review by Fan, Carilli, \& Keating
2006a). Studies of Gunn-Peterson (GP) absorption in quasar spectra
(Gunn \& Peterson 1965), and related phenomena, suggest a qualitative
change in the state of the intergalactic medium (IGM) at $z \sim 6$,
indicating a rapid increase in the neutral fraction of the IGM, from
$\rm f(HI) < 10^{-4}$ at $z \le 5.5$, to $\rm f(HI) > 10^{-3}$ (by
volume) at $z \ge 6$.  Conversely, the large scale polarization of the
cosmic microwave background (CMB) is consistent with instantaneous
reionization at $z \sim 10.4 \pm 1.2$ (Komatsu et al. 2010).  These
data suggest that reionization is less an event than a process,
beginning at $z \ge 10$, and with the `percolation', or 'bubble
overlap' phase ending at $z \sim 6$ to 7. Unfortunately, current
methods for measuring the neutral fraction of the IGM have fundamental
limitations, as GP absorption saturates at low neutral fractions ($\rm
f(HI) > 10^{-3}$), and CMB polarization measurements are an integral
measure of the Thompson scattering optical depth back to $z =
1000$. Indeed, Mesinger (2010) has pointed out that there is no
definitive evidence for (or against) a largely ionized IGM, even as
late as $z \sim 6$, while recent measurements of the temperature of
the high $z$ Ly$\alpha$ forest suggest fairly late reionization ($z <
9$; Bolton et al. 2010).

Luminous quasars at the end of cosmic reionization will generate the
largest ionized regions in the Universe during this epoch. These have
been alternatively called Cosmological Str\"omgren Spheres or quasar
near-zones (NZ), and we adopt the latter to be consistent with the
recent literature.  The near-zone behavior has been described by
numerous authors (Shapiro \& Giroux 1987; Madau \& Rees 2000; Cen \&
Haiman 2000). A straight-forward calculation (Haiman 2002) shows that,
when recombination is not important, and when the quasar lifetime is
much less than the age of the Universe, the physical size of the
expected NZ behaves as:

$$ \rm R_{NZ} = 8.0~ f(HI)^{-1/3}(\dot{N_Q}/6.5\times 10^{57} {\rm
s^{-1}})^{1/3} (t_Q/2\times 10^7 {\rm yr})^{1/3} [(1+{\it z}_Q)/7]^{-1} 
\rm Mpc $$

\noindent where $\rm R_{NZ}$ is the NZ proper radius, $\dot{\rm N_Q}$
is the rate of ionizing photons from the quasar, $z_Q$ is the quasar
host galaxy redshift\footnote{Note that we make the reasonable
assumption that the quasar is at rest relative to its host galaxy.},
and $\rm t_Q$ is the quasar age. This calculation ignores clumping,
and assumes the mean cosmic density over the large scales being
considered.

Spectra of $z \sim 6$ quasars show evidence for large ionized
regions around bright quasars on proper scales $\sim 5$ Mpc.  The
evidence comes in the form of excess transmission on the blue wing of
the Ly$\alpha$ emission line, prior to the onset of full GP
absorption. This excess emission has been interpreted as being due to
the local ionizing effect of the quasar.  A number of studies have
attempted to use these NZ to derive constraints on the IGM neutral
fraction, using the equation above and assuming mean quasar lifetimes
derived from demographics (White et al. 2003, 2005; Mesinger \& Haiman
2004; Wyithe, Loeb, \& Carilli 2005; Yu \& Lu
2005; Gnedin \& Prada 2004; Furlanetto, Hernquist \& Zaldariaga~2004;
Wyithe \& Loeb~2004; Walter et al. 2003).

The most detailed treatment of quasar near-zones associated with $z
\sim 6$ quasars to date is that by Fan et al. (2006b).  They point out
numerous factors that make the absolute measurement of the neutral
fraction using NZ radii difficult, including: (i) uncertainties in the
mean quasar lifetimes and ionizing photon rates, (ii) large scale
structure in biased regions and pre-ionization by local galaxies, and
(iii) clumpiness in the IGM (see also Lidz et al. 2007; Maselli et
al. 2007; Bolton \& Heahnelt 2007a,b).  However, Fan et al. also point
out that over the narrow redshift range being considered, the
systematics are likely to be comparable for all sources, and hence,
`if there is an order of magnitude evolution in the IGM ionization,
the size of the HII regions should show strong evolution, providing a
reliable {\sl relative} measurement of the neutral fraction (with
redshift).' 

More recently, Wyithe et al. (2008) and Bolton et al. (2010) have
argued, based on numerical simulations, that the dominant effect on
quasar near-zones during the end of reionization, or the percolation
stage, when the overall neutral fraction is low, may be a rapidly
increasing contribution to photo-ionization from the cosmic ionizing
background with cosmic time, acting in concert with the quasar photons
to lift the GP trough near the edge of the near-zone.

Fan et al. (2006b) emphasize that the measurements of the NZ sizes are
uncertain due to a number of factors, including the determination of
where the transmission drops to zero, ie. the on-set of the GP effect
($z_{GP}$), and the redshift of the host galaxy ($z_Q$). To mitigate
the former, they define the ionization zone, $\rm R_{NZ}$, as the
region in which the transmitted flux ratio (relative to the
extrapolated continuum) is above 0.1 when smoothed to a resolution of
20 $\AA$. This is well above the average GP transmission for quasars
at this redshift ($< 0.04$), and hence mitigates affects of a
fluctuating IGM, while still providing a good relative measurement of
NZ sizes over the narrow redshift range considered. Willott et al.
(2008) have considered this technique for two $z \sim 6$ quasars, and
point out that the transmission can rise above 0.1 just one or two
resolution elements further from the quasar, raising the question
of where to draw the line for $\rm R_{NZ}$?  In this
paper, we use the strict and quantifiable definition of Fan et
al. (2006b), namely, the first redshift below the quasar redshift at
which the transmission drops under 0.1, in order to be consistent with
previous literature.  We refer the reader to the extensive discussion
in Fan et al. 2006b and Willott et al. (2008) for details.

One major uncertainty in the Fan et al. (2006b) analysis was the host
galaxy redshift, $z_Q$.  In particular, most of the host galaxy
redshifts in the Fan et al. sample were based on a combination of
Ly$\alpha$+NV and high-ionization lines such as CIV and Si IV. It is
well documented that broad Ly$\alpha$ emission is a poor indicator of
the host galaxy redshift due to absorption on the blue-side of the
emission line, while the high ionization lines can have large velocity
offsets, of order 1000 km s$^{-1}$ ($\Delta z \sim 0.02$ at $z = 6$),
with respect to low ionization lines such as MgII (Richards et
al. 2002).

We have undertaken an extensive study of the radio through submm
properties of $z > 5.7$ quasars to probe the dust, molecular gas, and
star formation properties of the host galaxies (Wang et al. 2010; Wang
et al.  2008; Wang et al. 2007; Carilli et al. 2007; Walter et
al. 2009; Walter et al. 2004; Bertoldi et al.  2003; Bertoldi et
al. 2003).  One key result from our program is the detection of CO
emission from the host galaxies of 8 quasars (Wang et al.  2010). High
resolution imaging shows that the CO is centered within 0.2" of the
quasar, at least in one source (Walter et al. 2004; Riechers et
al. 2009), and CO almost certainly provides the host galaxy redshift
to very high accuracy ($\Delta z < 0.002$).  In parallel, we have been
using near-IR spectra to detect MgII emission from $z \sim 6$ quasars
(Kurk et al. 2007; Jiang et al. 2007), which also provides much more
accurate host galaxy redshifts ($\Delta z \le 0.007$).

\section{The quasars and redshifts}

We have assembled a sample of 27 $z\sim 6$ quasars, all of which have
reasonable signal-to-noise rest-frame UV spectra required for this
analysis, mostly from the Keck telescope.  Most of these (24) are from
the Sloan Digital Sky Survey from Fan et al. (2006c) and Jiang et
al. (2008).  J1425+3254 was discovered in the AGN and Galaxy Evolution
Survey of Cool et al. (2006). J2329-0301 and
J1509-1749 are from the Canada-France High-z Quasar Survey of Willott
et al. (2007).  Eight of this sample are detected in CO, from which an
accurate $z_Q$ is derived (see Wang et al. 2010 and references
therein).  Nine of the sources have $z_Q$ measured from the MgII
emission line (Kurk et al. 2007; Jiang et al. 2007). The other ten
objects have redshifts determined from the Ly$\alpha$+NV lines.

We summarize all the parameters in Table 1, including the quasar 1450
$\AA$ absolute magnitude ($\rm M_{1450}$), redshifts, and comoving
radial distances\footnote{Ned Wright's cosmological code is used in
the calculation (Wright 2006) with the standard WMAP cosmology
(Spergel et al. 2007).}.

We adopt the definition of the NZ radius ($\rm R_{NZ}$) used in Fan et
al. (2006b), i.e.  the physical distance implied by the Hubble flow,
derived from the difference between the quasar host galaxy redshift
($z_{\rm Q}$) to the point ($z_{\rm GP}$) where the transmitted flux
first drops by a statistically significant amount to below 
10\% of the quasar extrapolated continuum emission:
$\rm R_{p,NZ} = (D_Q - D_{GP})/(1+{\it z}_Q)$, after the spectrum is
smoothed to a resolution of $20\AA$.  $\rm D_Q$ and $\rm D_{GP}$ are
the co-moving distances implied by the redshifts.  Following Fan et
al. (2006b), we fit a double Gaussian + power-law profile to the
quasar spectrum redward of Ly$\alpha$ as the quasar intrinsic
continuum and then calculate transmitted flux as a function of
wavelength. 

Spectra of the SDSS quasars are from Fan et al. 2006 and Jiang et
al. 2008. The spectrum of the AGES quasar (J1425+3254) is from Cool et
al. (2006).  For the two Willott et al. quasars, J2329-0301 and
J1509-1749, we have not re-fit the spectra, but use the $\rm R_{p,NZ}$
values quoted in Willott et al. (2007). Again, to be consistent
with the sample in Table 1, we adopt the strict definition of Fan et
al. (2006) of $\rm R_{NZ}$ defined by the first redshift (below the
quasar redshift) at which the transmission drops below 0.1.

To properly compare NZ sizes, we normalize all the $\rm R_{p,NZ}$
values to a fixed UV absolute magnitude $\rm M_{1450}=-27$, using the
relation: $\rm
R_{NZ,corrected}=R_{p,NZ}\times10^{0.4(27+M_{1450})/3}$.  The $\rm
R_{NZ,corrected}$ values are listed in the last column of Table 1.

Three sources in Table 1, J0353+0104, J1044-0125, and J1048+4637, are
broad absorption line (BAL) quasars (Fan et al. 2006c; Jiang et
al. 2008), and the source J1335+3533 has a lineless UV spectrum
discovered in Fan et al. (2006c).  For completeness, we calculate the
$\rm R_{NZ}$ values for these quasars in Table 1. However, as with
Fan et al. (2006b), we exclude these quasars from our subsequent
analysis due to the fundamentally different nature of their intrinsic
spectra, in particular around the Ly$\alpha$ emission redshift. The
different intrinsic spectra imply that using the same spectral
fitting analysis technique described above to derive $z_{\rm GP}$
could lead to erroneous results.  Keeping this cautionary note in
mind, it is interesting that these sources are all among the smallest
near-zones in the sample, and three of the four have been detected in
CO. The presence of a BAL, and of CO in the host galaxy, implies
relatively dense local environments for these quasars, and hence
larger UV attenuation, which could lead to small $\rm R_{NZ}$ values.

\section{Redshift errors}

The Ly$\alpha$+NV redshifts have large
uncertainties of $\rm \Delta z \sim 0.02$, due to emission broad line
widths, Gunn-Peterson absorption, and offsets between different
ionization lines in quasar spectra (Richards et al. 2002, 2003; Fan et
al. 2006c). This implies an uncertainty of $\Delta\rm R_{NZ} \sim 1.2$
Mpc. There is also a systematic offset of similar magnitude between
the MgII and higher ionization broad line redshifts (Richards et
al. 2002).  We have corrected the redshifts in the bottom section of
Table 1 ('other redshifts') using this mean offset ($+0.02$ in $z$;
note that J2315-0023 was already corrected in the literature).

Richards et al. (2002) compare MgII broad line redshifts with [OIII]
narrow line redshifts in lower $z$ quasars and find: $z(MgII) -
z([OIII]) = -97 \pm 269$ km s$^{-1}$.  We add to this the
uncertainties in spectral fitting ($\Delta z \sim 0.003$; Jiang et
al. 2007; Kurk et al. 2007), and estimate a typical uncertainties in
the MgII redshifts of $\Delta z \sim 0.007$, implying $\Delta\rm
R_{NZ} \sim 0.4$ Mpc. The CO observations significantly improve the
redshift measurements, with an uncertainty of $\Delta z \le 0.002$,
implying $\Delta\rm R_{NZ} \sim 0.1\,Mpc$.

Considering the errors on $z_{\rm GP}$, Fan et al. (2006b) estimate
that at $20\AA$ resolution, they can measure the onset of the
Gunn-Peterson trough to an accuracy of at least $10\AA$, or $\Delta
z_{\rm GP} \sim 0.01$. This is certainly a conservative estimate of
the error, and implies $\rm \Delta\rm R_{NZ} \sim 0.6$ Mpc due to errors
in $z_{\rm GP}$. We add this error in quadrature to the errors on
$z_Q$. The implied total errors are then: (i) $\rm \Delta R_{NZ} =
1.34$Mpc (physical) for sources with Ly$\alpha$ redshifts, (ii)
$\rm \Delta R_{NZ} = 0.72$Mpc for sources with MgII redshifts, and
(iii)$\rm \Delta R_{NZ} = 0.60$Mpc for sources with CO redshifts.

\section{Analysis} 

We plot the $\rm R_{NZ,corrected}$ values versus quasar redshift in
Figure 1. Again, we omit the three BAL sources and the lineless quasar
from this analysis for reasons given in Section 2.  The figure shows a
decrease in $\rm R_{NZ}$ with increasing redshift.  A weighted linear
fit to the data points gives $\rm R_{NZ,corrected} = (7.4\pm 0.3) -
(8.0 \pm 1.1) \times (z-6)$, while an unweighted fit yields a similar
result of $\rm R_{NZ,corrected} = (7.7\pm 0.4) - (6.8 \pm 2.1) \times
(z-6)$.  The errors were computed both from the formal fitting
process, as well as by computing the scatter in the fitted parameters
after fitting realizations removing one data point at time. To be
conservative, we adopt the larger errors from the latter process in
the analysis below. 

For the weighted fit, the mean NZ radius changes from $9.8 \pm 0.9$
Mpc at $z = 5.7$, to $4.2\pm 0.4$ Mpc at $z = 6.4$.  The Spearman rank
correlation coefficient for the distribution is $\rho = -0.51$,
implying a probability that there is no correlation between
redshift and $\rm R_{NZ,corrected}$ of just 1\%. Further, a Student's
t-test gives a probability of 1\% for the hypothesis that the slope is
zero (ie. no change in mean NZ size over this redshift range).  In the
simplest physical picture, $\rm R_{NZ} \propto$
f(HI)$^{-1/3}(1+z_Q)^{-1}$, implying an increase in the volume
averaged neutral fraction by a factor $\sim 9.4$ from $z = 5.7$ to
6.4.

After removing the linear slope in redshift, we calculate an rms for
the scatter in the normalized NZ sizes of $\rm \sigma_{NZ} = 2$ Mpc.
Fan et al. (2006b) see a similar scatter, but could not say whether
this was due to errors in $z_Q$ or physical differences between
quasars. In our sample, the measurement errors are all below this
rms value, and hence there must be intrinsic physical differences
between quasars, either in terms of their environments, or AGN
properties (eg. quasar age, SED).

Figure 2 shows the mean absorption profiles in the quasar NZ for
three different redshift bins. This is the updated version of Fig. 13
in Fan et al. (2006b), and as with Fan et al., we do not scale the
profile according to the quasar luminosity in this case.  There is a
clear evolution in profiles. The lowest redshift bin first drops below
10\% transimission by a statistically significant amount at 8.5 Mpc
radius, while this occurs for the highest redshift bin at 3.5 Mpc.
Fan et al. (2006b) point out that this marked difference in absorption
profiles from $z \sim 5.85$ to $z > 6.15$ is a graphic illustration of
the change in the mean neutral state of the IGM at $z \sim 6$.

Figure 3 shows the relationship between $\rm R_{p,NZ}$ and
$M_{1450}$. Figure 3a shows $\rm R_{p,NZ}$ values taken directly from
Table 1, and Figure 3b shows the values after normalizing-out the
linear regression relationship between NZ size and redshift
derived from Figure 1, which we designate as $\rm R_{p,NZ,norm}$.  A
correlation is found between NZ sizes and $M_{1450}$.
The Spearman rank correlation coefficient for the distribution in
Figure 3b is $\rho = -0.57$, implying a probability that there
is no correlation between M$_{1450}$ and $\rm R_{p,NZ,norm}$ of just
0.4\%.  Included in Figure 3 is a dotted line that follows the
relationship: $\rm R_{\rm p,NZ} \propto \dot{N}_Q^{1/3}$, as naively
expected for photo-ionization dominated by quasar radiation. This line
has not been fit to the data, but is shown for reference. The overall
trend of increasing $\rm R_{p,NZ}$ with the quasar luminosity is
consistent with this very simple model, although the scatter is such
that this is by no means proof of the model.

As a final comparison, Figure 4 shows the relationship between
$M_{1450}$ and redshift.  We find no correlation between quasar
absolute magnitude and redshift over the narrow redshift range of
this sample. Formally, the Spearman rank
correlation coefficient for the distribution is $\rho = 0.11$,
implying a high probability that there is no correlation between
redshift and M$_{1450}$ of 62\%.

\section{Discussion}

Using a larger sample and improved quasar host redshifts, we have
confirmed a statistically significant increase in luminosity
normalized NZ sizes from $z \sim 6.4$ to 5.7, first noted by Fan et
al. (2006b).  In the simplest physical model, the factor 2.3 change
in mean NZ radius is consistent with a factor of 9.4 decrease in mean
neutral fraction from  $z \sim 6.4$ to 5.7. Fan et al. (2006b) derive a
volume averaged neutral fraction of $9.3\times 10^{-5}$ at $z = 5.7$
from their analysis of GP absorption troughs and gaps.  The NZ
measurements then suggest that the neutral fraction has increased to
$\sim 9\times 10^{-4}$ at $z = 6.4$.

Wyithe et al. (2008) and Bolton et al. (2010) have proposed an
alternative interpretation for this evolution in NZ sizes. Based on
numerical simulations, they argue that the dominant effect on quasar
near-zones during the end of reionization, or the percolation stage,
may be a rapidly increasing contribution to photo-ionization from the
cosmic ionizing background with cosmic time. The increasing mean free
path allows the background photons to contribute substantially to the
ionization around the edge of the NZ, thereby effectively increasing
the observed sizes of the NZ.  In either case, the results are
consistent with the idea that $z \sim 6$ to 7 corresponds to the tail
end of cosmic reionization.

Our improved $z_Q$ measurements imply that the intrinsic scatter in
the NZ sizes is large at any given redshift: $\rm \sigma_{NZ} = 2$
Mpc. This large scatter must reflect real physical differences between
quasars, perhaps some combination of differences in SEDs, UV escape
fractions, ages, and/or local environments.  Bolton \& Haehnelt
(2007b) have argued that the quasar age is less a factor when f(HI) $<
0.1$, since the size is then most likely dictated by absorption (and
recombination) in denser regions of residual neutral gas within the
sphere itself. The small NZ sizes seen for the BAL quasars in Table 1
suggest a dependence on the immediate local environment, although see
the cautionary note in Section 2 concerning these sources.

We also find an increase in $\rm R_{p,NZ}$ with quasar UV luminosity.
This correlation is qualitatively consistent with photo-ionization
dominated by quasar radiation.  Lastly, we find no correlation between
$\rm M_{1450}$ and redshift over the relatively narrow redshift range
explored in this sample.

We re-emphasize that the use of the near-zone technique has major
pit-falls when trying to estimate the absolute neutral fraction at a
given redshift, as detailed in Fan et al. (2006b) and Bolton \&
Heahnelt (2007a,b). In this paper, we have improved substantially one
of the observational uncertainties in the calculation, namely, the
quasar host galaxy redshifts, and we have focused on the evolution of
the NZ sizes and neutral fraction, not the absolute neutral
fraction. Still, as Bolton \& Heahnelt point out, a larger sample with
accurate measurements of the spectral regions around both Ly$\alpha$
and Ly$\beta$ is necessary to set meaningful constraints on the
absolute neutral fraction of the IGM using the NZ technique.

\acknowledgements 
CC thanks the Max-Planck-Gesellschaft and the Humboldt-Stiftung for
support through the Max-Planck-Forschungspreis, and the Max-Planck
Institute for Astronomie in Heidelberg for their hospitality.  The
authors thank J. Bolton, S. Wyithe, C. Willott, and the referee for useful
comments. DR acknowledges support from NASA through Hubble
Fellowship grant HST-HF-51235.01 awarded by the STScI, operated by
AURA for NASA, under contract NAS 5-26555.

\clearpage
\newpage

\noindent  Bertoldi, F., Cox, P., Neri, R. 2003, A\& AL, 409, L47

\noindent  Bertoldi, F., Carilli, C., Cox, P. 2003, A\& AL, 406, L55

\noindent Bolton, J. \& Haehnelt, M. 2007a, MNRAS, 374, 493

\noindent Bolton, J. \& Haehnelt, M. 2007b, MNRAS, 381, L35

\noindent Bolton, J., Becker, G., Wyithe, J.S. et al. 2010, 
MNRAS, in press (arXiv:1001.3415)

\noindent  Carilli, C., Neri, R., Wang, R. et al. 2007, ApJL, 666, L9

\noindent  
Cen, R. \& Haiman, Z., 2000, ApJ, 542, L74

\noindent  Fan, X., Carilli, C., Keating, B. 2006a, ARAA, 44, 415

\noindent  
Fan, X, Strauss, M., Becker, R. et al. 2006b, AJ, 132, 117

\noindent  
Fan, X, Strauss, M., Richards, G. et al. 2006c, AJ, 131, 1203

\noindent  
Fan, X, Strauss, M., Becker, R. et al. 2002, 123, 1247

\noindent  
Fan, X, et al. 2003, AJ, 125, 1649

\noindent  
Furlanetto, S.R. , Hernquist, L. \& Zaldarriaga, M. 2004, MNRAS, 354, 695

\noindent  Gnedin, N.~Y.\ 2004, \apj, 610, 9

\noindent 
Gnedin, N.Y., Prada, F. 2004, ApJ, 608, L77

\noindent  
Gunn, J.~E.~\& Peterson, B.~A.\ 1965, ApJ, 142, 1633

\noindent Haiman, Z.\ 2002, \apjl,
576, L1

\noindent  Jiang, L., Fan, X., Vestergaard, M.  et al. 2007, AJ, 134, 1150

\noindent  Komatsu, E., Smith, K., Dunkley, J. et al. 2010,
ApJS, in press (arXiv:1001.4538)

\noindent  Kurk, J., Walter, F., Fan, X. et al. 2007, ApJ, 669, 32

\noindent  Lidz, A., McQuinn, M., Zaldarriaga, M., Hernquist, 
L., Dutta, S. 2007, ApJ, 670, L39

\noindent  
Madau, P., Rees, M.J., 2000, ApJ, 542, L69

\noindent  Maselli, A., Gallerani, S., Ferrara, A., Choudhury, T.
2007, MNRAS, 376, L34

\noindent 
Mesinger, A. \& Haiman, Z. 2004, ApJ, 611, L69

\noindent  Mesinger, A. 2010, MNRAS, in press (arXiv:0910.4161)



\noindent  
Richards, G.T., et al., 2002, AJ, 124, 1

\noindent  --------------------. 2003, Astron.J., 127, 1305

\noindent  Riechers, D., Walter, F., Bertoldi, F. et al. 2009,
ApJ, 703, 1338

\noindent Shapiro, P. \& Giroux, M. 1987, ApJL, 321, 107

\noindent  Spergel, D., Bean, R., Dore, O., et al. 2007, ApJS, 170, 377

\noindent 
Walter, F., Bertoldi, F., Carilli, C.L. et al., 2003, Nature, 424, 406

\noindent 
Walter, F., Carilli, C., Bertoldi, F., Menten, K., Cox, P., Lo, K.Y., 
Fan, X., Strauss, M., 2004, ApJ, 615, L17

\noindent  Walter, F., Riechers, D., Cox, P. et al. 2009, Nature,
457, 699

\noindent Wang, R., Carilli, C.L., Wagg, J. et al. 2010, ApJ, submitted

\noindent Wang, R., Carilli, C.L., Wagg, J. et al. 2008, ApJ, 687, 848

\noindent Wang, R., Carilli, C.L., Beelen, A. et al. 2007, AJ, 134, 617

\noindent  White, R., Becker, R., Fan, X., Strauss, M. 2003, {AJ}, {126}, 1 

\noindent  White, R., Becker, R., Fan, X., Strauss, M. 2005, AJ, 129, 2102

\noindent  Willott, C.J., Delorme, P., Reyle, C. et al. 2009,
arXiv:0912.0281

\noindent  Willott, C.J., Delorme, P., Omont, A., et al. 2007,
AJ, 134, 2435

\noindent  Wright, E. 2006, PASP, 118, 1711

\noindent  
Wyithe, J.~S.~B.~\& Loeb, A.\ 2004, ApJ, 612, 597

\noindent  Wyithe, J.S., Loeb, A., Carilli, C. 2005, ApJ, 628, 575

\noindent  Wyithe, J.S., Bolton, J., Haehnelt, M. 2008, MNRAS,
383, 691

\noindent  
Yu, Q., Lu, Y. 2004, ApJ, 602, 603

\clearpage
\newpage

\begin{table}
{\scriptsize
\caption{Redshifts, comoving distances, and proper NZ radii}
\begin{tabular}{lccccccc}
\hline \noalign{\smallskip} \hline \noalign{\smallskip}
\multicolumn{8}{c}{CO redshifts$^a$}\\
\noalign{\smallskip} \hline \noalign{\smallskip}
Name & $\rm M_{1450}$ & $z_Q$ & $\rm D_Q$ & $\rm z_{GP}^b$ & $\rm D_{GP}$ & 
$\rm R_{\rm p,NZ}^c$ & $\rm R_{NZ,corrected}$ \\
     &                &                  &  Mpc             &               & 
Mpc           & Mpc  & Mpc \\
\noalign{\smallskip} \hline \noalign{\smallskip}
J084035.0+562419 & -26.66 & 5.8441 & 8352.6 & 5.69 & 8281.8 & 10.4 & 11.5 \\
J092721.8+200123 & -26.78 & 5.7722 & 8319.9 & 5.70 & 8286.5 & 4.9 & 5.3  \\
J104433.0$-$012502$^d$ & -27.47 & 5.7824 & 8324.5 & 5.70 & 8286.5 & 5.7 & 4.9 \\
J104845.0+463718$^d$ & -27.55 & 6.2284 & 8519.2 & 6.16 & 8490.5 & 4.0 & 3.4 \\
J114816.6+525150 & -27.82 & 6.4189 & 8596.9 & 6.33 & 8561.0 & 4.9 & 3.8 \\
J133550.8+353315$^d$ & -26.82 & 5.9012 & 8378.2 & 5.89 & 8373.2 & 0.76 & 0.8 \\
J142516.3+325409 & -26.09 & 5.8918 & 8374.0 & 5.76 & 8314.3 & 8.8 & 11.6 \\
J205406.4$-$000514 & -26.15 & 6.0379 & 8438.3 & 5.97 & 8408.7 & 4.2 & 5.5 \\
\noalign{\smallskip} \hline \noalign{\smallskip}
\multicolumn{7}{c}{Mg II redshifts$^e$}\\
\noalign{\smallskip} \hline \noalign{\smallskip}
J000552.3$-$000655 & -25.87 & 5.850 & 8355.2 & 5.77 & 8318.9 & 5.3 & 7.5 \\
J030331.4$-$001912 & -25.48 & 6.078 & 8455.6 & 6.00 & 8422.2 & 4.7 & 7.5  \\
J083643.8+005453 & -27.88 & 5.810 & 8337.1 & 5.62 & 8248.9 & 13.0 & 9.9 \\
J103027.1+052455 & -27.16 & 6.308 & 8552.0 & 6.21 & 8511.5 & 5.6 & 5.3 \\
J130608.2+035626 & -27.19 & 6.016 & 8428.8 & 5.92 & 8386.6 & 6.0 & 5.7 \\
J141111.2+121737 & -26.74 & 5.904 & 8379.5 & 5.82 & 8341.7 & 5.4 & 5.9 \\
J150941.7$-$174926 & -26.98 & 6.118 & --    & -- & -- &  4.3  & 4.3 \\
J162331.8+311200 & -26.67 & 6.247 & 8526.9 & 6.16 & 8490.5 & 5.1 & 5.6 \\
J163033.9+401209 & -26.12 & 6.065 & 8450.0 & 5.94 & 8395.4 & 7.7 & 10.1 \\
\noalign{\smallskip} \hline \noalign{\smallskip}
\multicolumn{7}{c}{Other Redshifts$^f$}\\
\noalign{\smallskip} \hline \noalign{\smallskip}
J000239.3+255034 & -27.67 & 5.82 & 8341.7 & 5.65 & 8263.1 & 11.5 & 9.4 \\
J035349.7+010404$^d$ & -26.53 & 6.049 & 8443.1 & 6.02 & 8430.5 & 1.8 & 2.1 \\  
J081827.4+172251 & -27.40 & 6.02 & 8430.5 & 5.89 & 8373.2  & 8.1 & 7.2 \\
J084119.5+290504 & -27.12 & 5.98 & 8413.0 & 5.81 & 8337.1  & 10.9 & 10.5 \\
J113717.7+354956 & -27.12 & 6.03 & 8434.9 & 5.91 & 8382.1 & 7.5 & 7.2  \\
J125051.9+313021 & -27.14 & 6.15 & 8486.3 & 6.03 & 8434.9 & 7.2 & 6.9  \\
J143611.7+500706 & -26.54 & 5.85 & 8355.2 & 5.72 & 8295.8 & 8.7 & 10.0  \\
J160253.9+422824 & -26.83 & 6.09 & 8460.7 & 5.94 & 8395.4 & 9.2 & 9.7  \\
J231546.5$-$002358 & -25.43 & 6.117 & 8472.3 & 6.05 & 8443.5 & 4.1 & 6.6 \\
J232908.2$-$030158 & -25.23 & 6.427 & -- & -- & -- & 3.7 & 6.4 \\
\noalign{\smallskip} \hline
\end{tabular}
}

{\scriptsize
~$^a$Errors on $z_Q$ based on CO redshifts are $\Delta z \sim 0.002$. \\
~$^b$Errors on $z_{\rm GP}$ at 20$\AA$ resolution 
estimated by Fan et al. (2006b)  are $\Delta z \sim 0.01$. \\
~$^c$In section 3, we estimate errors in $\rm R_{p,NZ}$ as: 
$\rm \Delta R_{p,NZ} =$ 1.34 Mpc for sources with Ly$\alpha$ redshifts, 
0.72 Mpc for sources with MgII  redshifts, and 0.60 Mpc for sources with CO redshifts. \\
~$^d$BAL or lineless quarsar, see Section 2. \\
~$^e$Errors on $z_Q$ based on MgII redshifts are $\Delta z \sim 0.007$. \\
~$^f$Errors on $z_Q$ based on Ly$\alpha$+NV redshifts are  $\Delta z \sim 0.02$. \\}
\end{table}

\begin{figure}
\epsscale{1.0}
\plotone{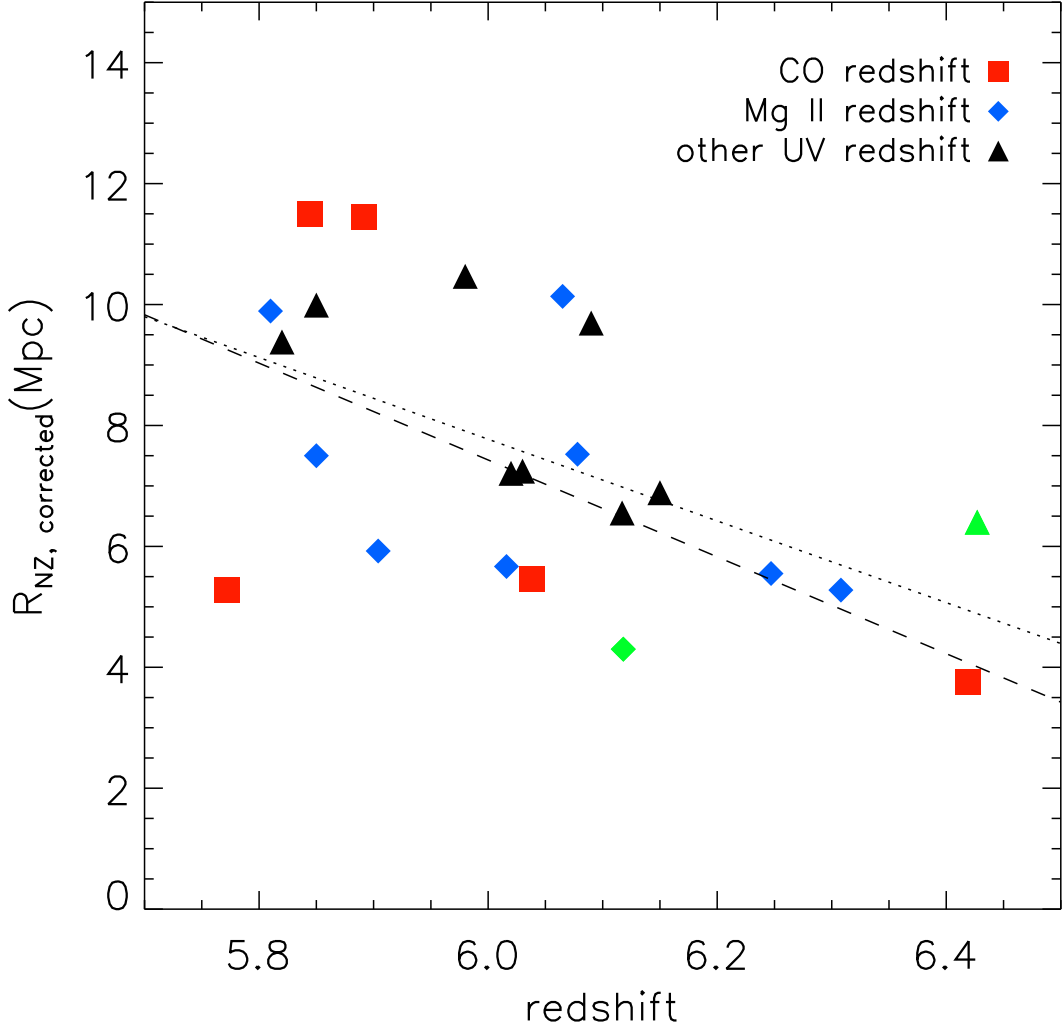}
\caption{
Radii of the luminosity normalized cosmological Str\"omgren spheres ($\rm
R_{NZ,corrected}$) around the z$\sim$6 quasars versus their
redshifts. Data are from Table 1, and again, 
we omit the three BAL sources and the lineless quasar
from this plot and analysis for reasons given in Section 2.
The red squares represent the  sources which have CO
redshifts. The blue diamonds represent sources with redshifts
determined with Mg II $\rm \lambda$ 2798$\rm \AA$ line emission, and
the black triangles denote sources with redshifts determined with
other UV lines from the discovery papers, such
as Ly$\alpha$ + NV (see Fan et al. 2006b,c and references
therein; Jiang et al. 2008). 
The errors for the different data points are: 
$\rm \Delta R_{NZ,corrected} =$ 1.34 Mpc, 0.72 Mpc, and 0.60 Mpc, 
for the UV, MgII, and CO-determined $z_Q$ 
uncertainties, respectively, including the error in $z_{\rm GP}$
added in quadrature. The long dash line show a weighted linear fit to the
data with $\rm R_{NZ,corrected} = (7.4 \pm 0.3) - (8.0 \pm 1.1) \times
(z-6)$. The dotted line shows an unwieghted fit. 
The two quasars from Willott et al. (2008) are shown in green,
with the $\rm R_{NZ,corrected}$ taken directly from that paper. 
}
\end{figure}

\begin{figure}
\epsscale{1.0}
\plotone{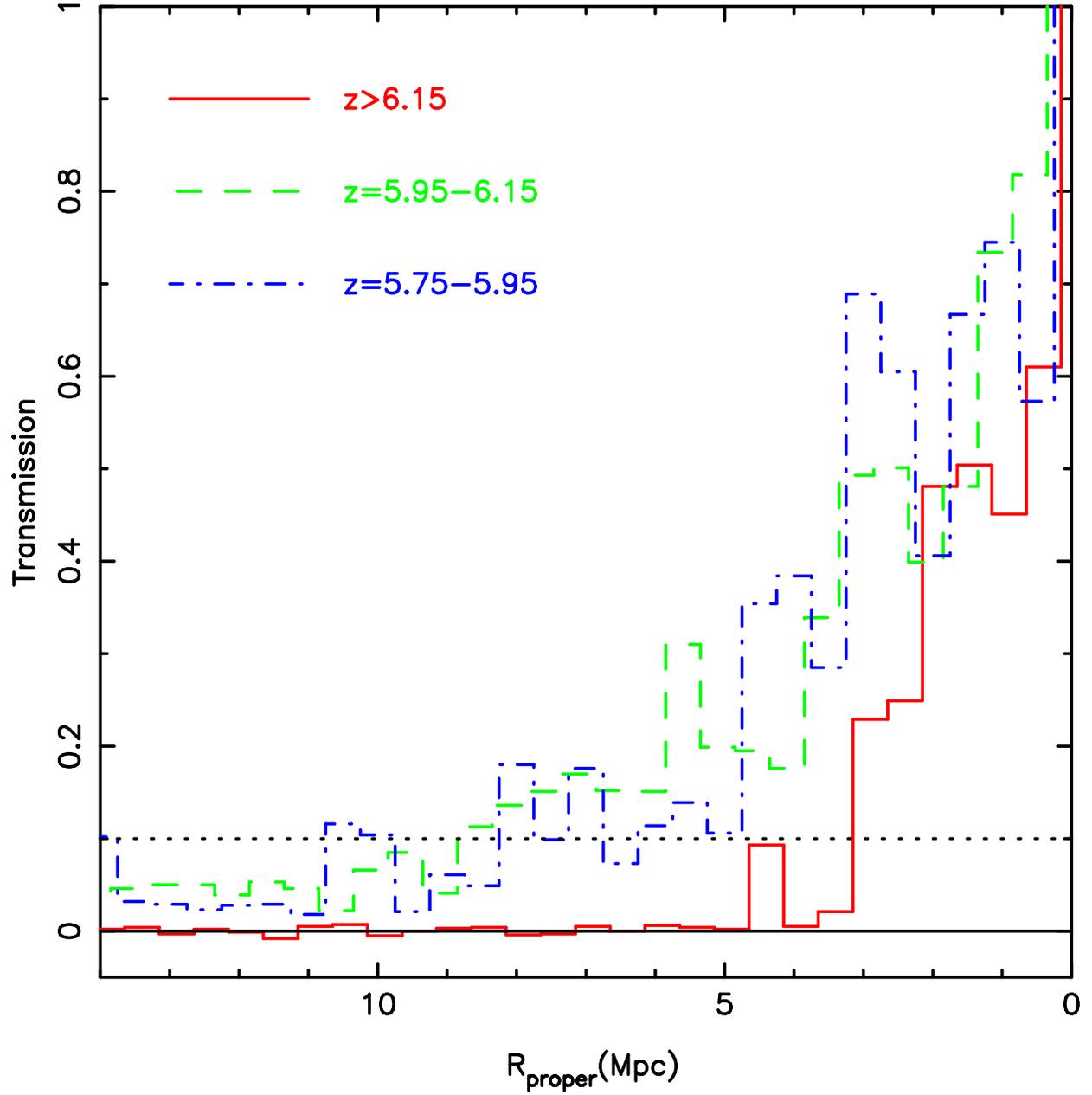}
\caption{
Average absorption profiles in the quasar near-zone for quasars
in three different redshift bins: $z=5.75 - 5.95$, $z=5.95 - 6.15$ and
$z>6.15$. The Ly$\alpha$+NV emission profile has been 
fitted with a Gaussian and removed. Note that the two Willott et al. 
quasars are not included in this plot. 
}
\end{figure}

\begin{figure}
\epsscale{0.6}
\plotone{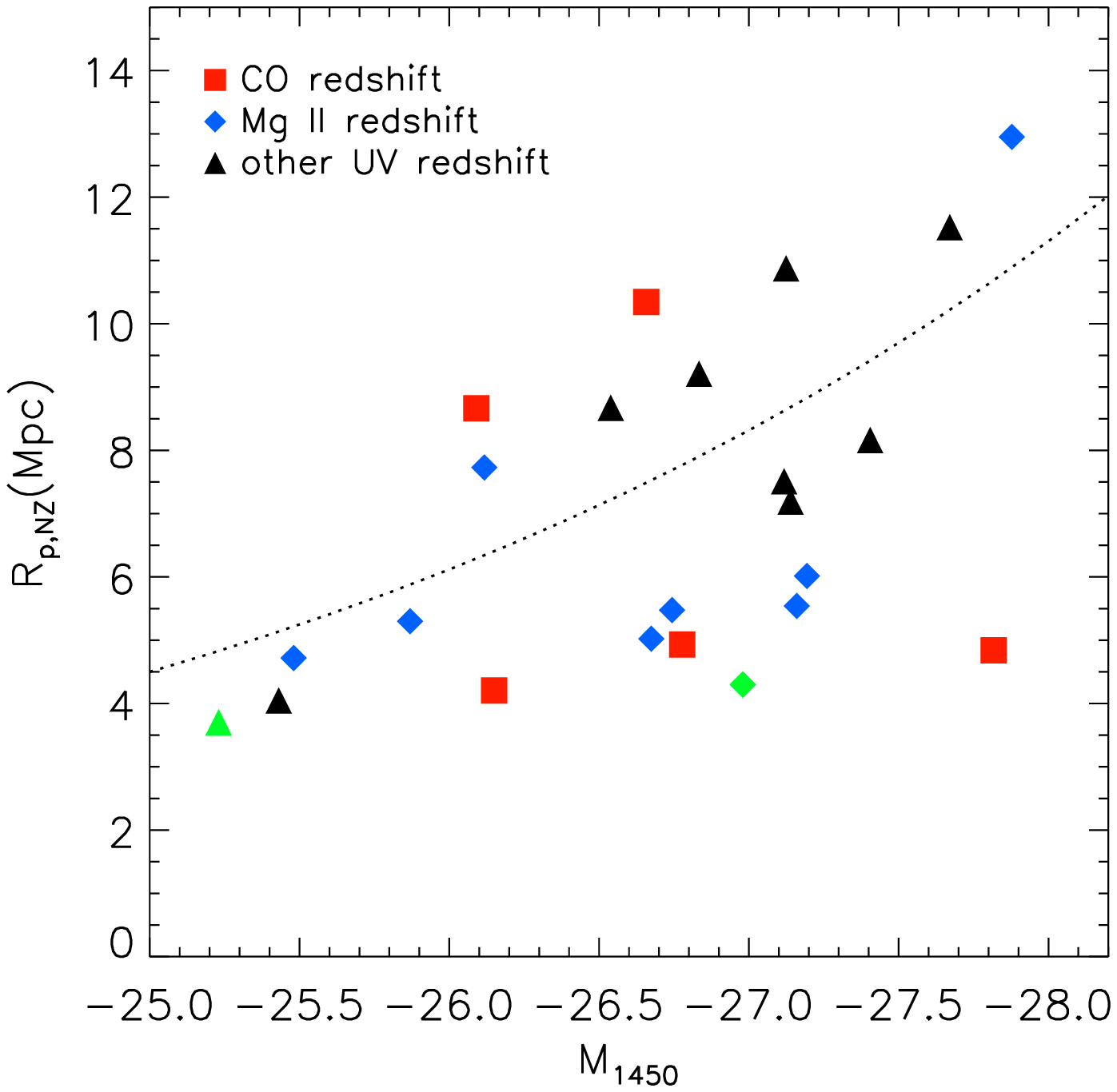}
\plotone{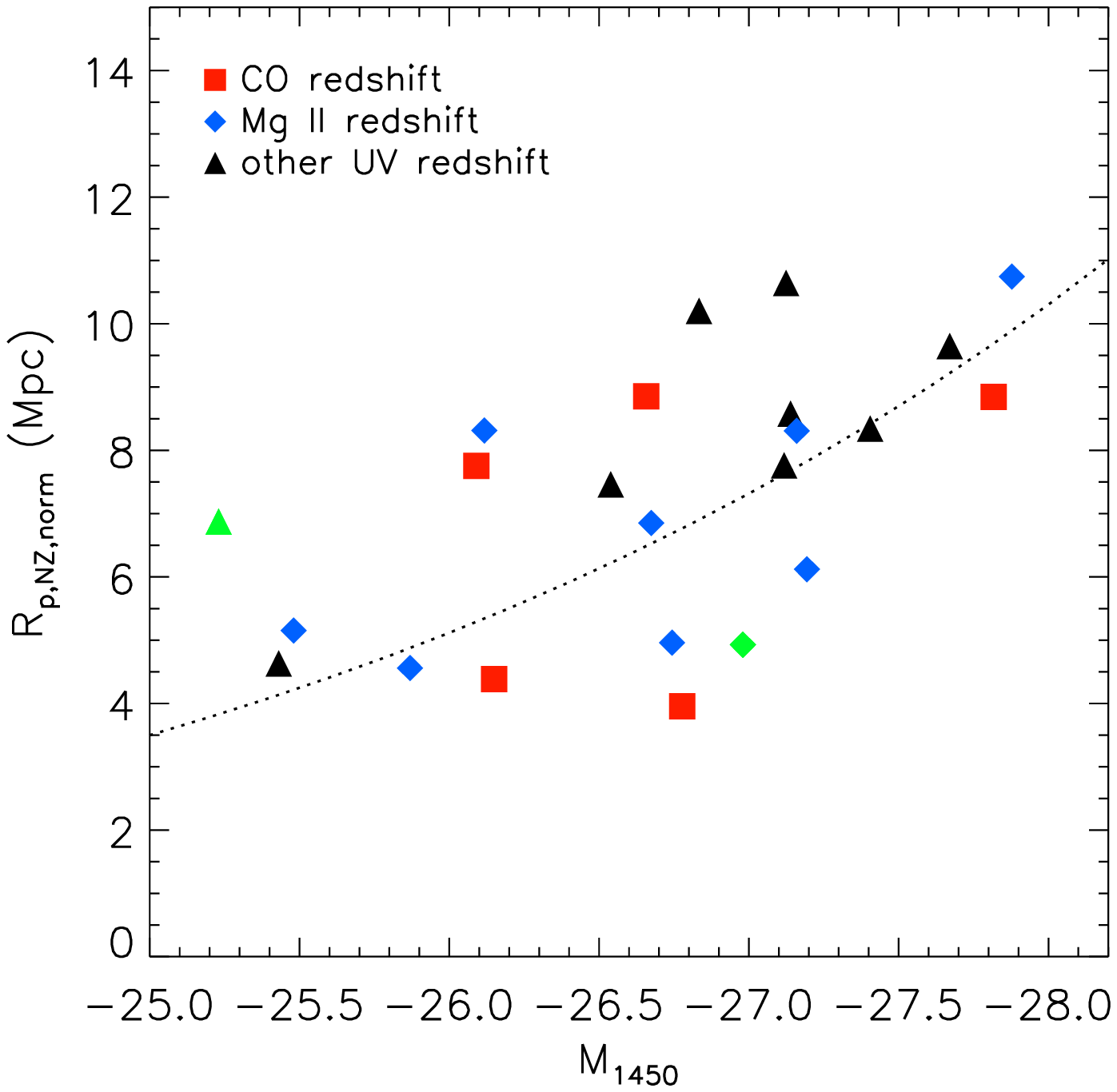}
\caption{
The relationship between $\rm R_{\rm p,NZ}$ and $M_{1450}$. 
The upper panel shows the values of  $\rm R_{\rm p,NZ}$ as given
in Table 1. The lower panel shows these same values with the
linear relationship between redshift and  NZ size 
derived from Figure 1 normalized-out. 
The dotted line is not a fit to the data, but shows the
relationship: $\rm R_{NZ} \propto \dot{N}_Q^{1/3} \propto 10^{-M_{1450}/7.5}$, 
as expected for photo-ionization dominated by quasar radiation. The
scaling for this curve is arbitrary. The symbols are the same as in Figure 1. 
}
\end{figure}

\begin{figure}
\epsscale{1.0}
\plotone{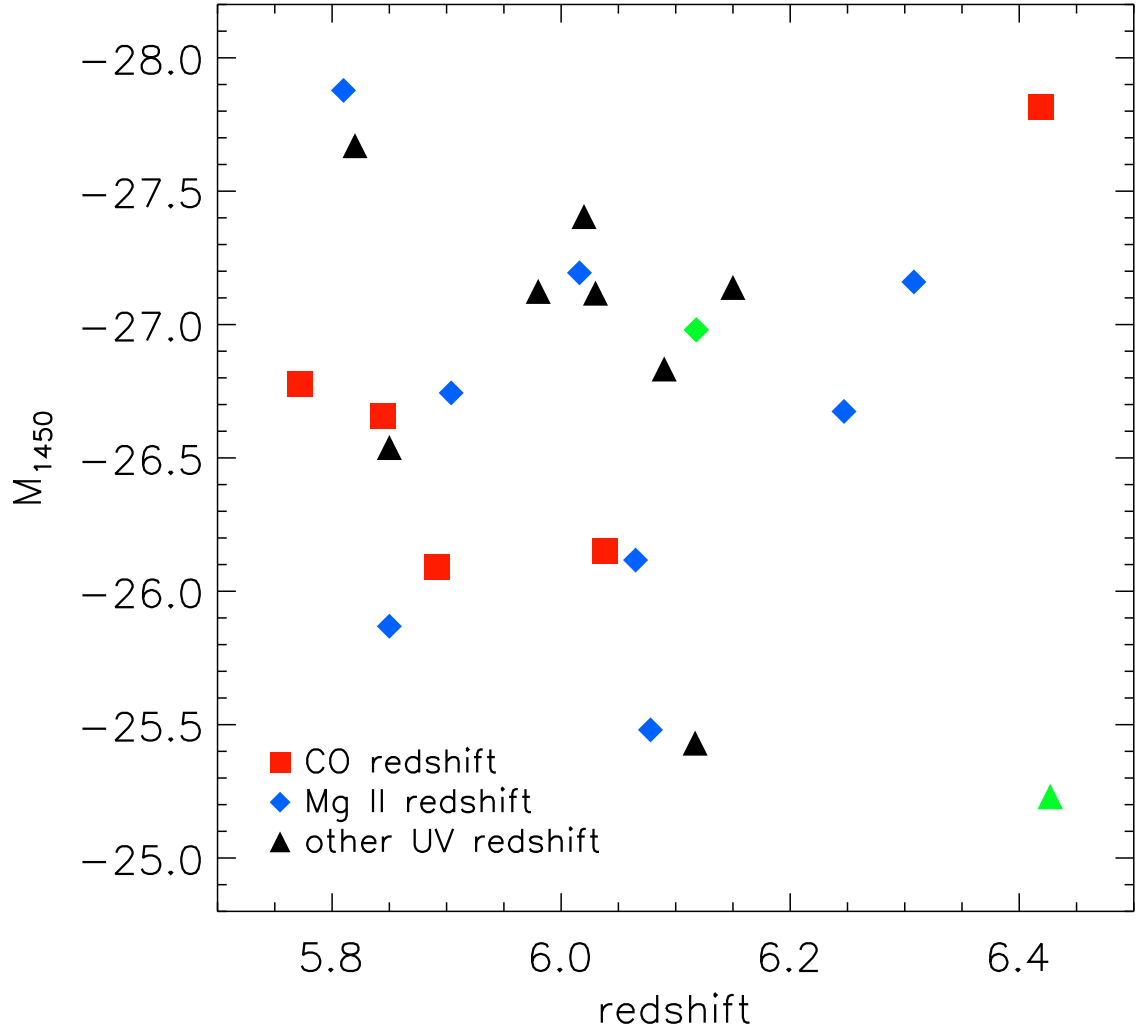}
\caption{
The relationship between redshift and $M_{1450}$.  
The symbols are the same as in Figure 1. 
}
\end{figure}

\end{document}